\documentclass{pasj00}
\draft

\begin{document}
\SetRunningHead{M.Yamagishi et al.}{AKARI infrared observations of edge-on spiral galaxy NGC~3079}

\title{AKARI infrared observations of edge-on spiral galaxy NGC~3079}

\author{Mitsuyoshi \textsc{Yamagishi}, Hidehiro \textsc{Kaneda}, Daisuke \textsc{Ishihara}}
\affil{Graduate School of Science, Nagoya University, Chikusa-ku, Nagoya 464-8602}
\email{yamagishi@u.phys.nagoya-u.ac.jp, kaneda@u.phys.nagoya-u.ac.jp, ishihara@u.phys.nagoya-u.ac.jp,}
\author{Shinya {\sc Komugi}}
\affil{Institute of Space and Astronautical Science, Japan Aerospace Exploration Agency, 3-1-1 Yoshinodai, Sagamihara, Kanagawa 229-8510}
\email{skomugi@ir.isas.jaxa.jp}
\author{Takashi {\sc Onaka}}
\affil{Department of Astronomy, Graduate School of Science, The University of Tokyo, 7-3-1 Hongo, Bunkyo-ku, Tokyo 113-0033}
\email{onaka@astron.s.u-tokyo.ac.jp}
\and
\author{Toyoaki {\sc Suzuki}}
\affil{Advanced Technology Center, National Astronomical Observatory of Japan, Mitaka, Tokyo 181-8588}
\email{suzuki@ir.isas.jaxa.jp}


%


\maketitle

\begin{abstract}
We present AKARI near- to far-infrared images of the nearby edge-on spiral galaxy NGC~3079 in 10 photometric bands.
The spectral energy distribution consists of continuum emission from dust with a single temperature of 28--33 K together with strong mid-infrared emission features from polycyclic aromatic hydrocarbons (PAHs). 
We derive the dust masses of $5.6 \times 10^6 \MO$ and $1.4 \times 10^7 \MO$ for the central 4 kpc region and the whole galaxy, respectively, and find that a gas-to-dust mass ratio is unusually high in the central region ($\sim$ 1100) and even for the whole galaxy ($\sim$ 860).
The ratio of the surface brightness distribution at the wavelength of 7 $\mathrm{\mu m}$ to that at 11 $\mathrm{\mu m}$ suggests that the properties of PAHs have spatial variations. Emission from ionized and neutral PAHs is relatively strong in the center and the disk regions, respectively, suggesting stronger radiation field and thus relatively active star formation in the center. Yet the total infrared luminosities of the galaxy indicate rather low star formation rates.
These results suggest that NGC~3079 is in an early-phase starburst stage.
\end{abstract}

\section{Introduction}

NGC~3079 is a highly-inclined ($i=84^\circ$) edge-on spiral galaxy at a distance of 15.6 Mpc, the optical disk of which has a position angle of $166^\circ$ (Irwin \& Seaquist 1991; Sofue \& Irwin 1992).
The nucleus of NGC~3079 is classified as a low-ionization nuclear emission-line region (LINER; Heckman 1980) or Seyfert 2 (Ford et al. 1986).
The galaxy is known to have an ultra-high-density molecular core ($3 \times 10^8 \MO$) in the center (Sofue et al. 2001).
In addition, molecular gas with a total mass of $5 \times 10^9 \MO$ is contained within the central 2 kpc radius (Koda et al. 2002).
The nucleus shows very bright $\mathrm{H_2 O}$ emission, which is one of the most luminous $\mathrm{H_2 O}$ maser sources (Henkel et al. 1984).
Near the central region, kilo-parsec-scale outflows along the minor axis are observed at various wavelengths (radio continuum: Duric \& Seaquist 1988; H$\alpha$: Ford et al. 1986; Cecil et al. 2001; X-ray: Fabbiano et al. 1992).
The superbubbles may have been produced by either a central active galactic nucleus (AGN, i.e. acretion of dense gas into the core; Irwin \& Seaquist 1988; Hawarden et al. 1995) or a nuclear starburst (Sofue \& Vogler 2001), but the exact origin is yet to be understood.

Infrared (IR) imaging observations of NGC~3079 are relatively few, in contrast to other wavelengths, mainly due to low spatial resolution especially in the far-IR. 
IR observations however provide very important tools to study dust components that trace star formation activities effectively.
Young et al. (1989) and Klaas \& Walker (2002) discussed the physical states of far-IR dust in NGC~3079 from the observations of IRAS and ISO, respectively.
However the angular resolutions of IRAS and ISO are so large as compared to the apparent size of NGC~3079 ($\timeform{7'.9} \times \timeform{1'.4}$; de Vaucouleurs et al. 1991) and therefore IRAS and ISO could not observe the galaxy as a spatially resolved far-IR source.
Klaas \& Walker (2002) derived the dust-to-gas mass ratio of 200 for the whole galaxy of NGC~3079, which is close to the accepted value of 100--200 for our Galaxy (Sodroski et al. 1997).
Braine et al. (1997) and Stevens \& Gear (2000) discussed the properties of dust in NGC~3079 on the basis of the spatially resolved submillimeter dust emission with IRAM and SCUBA, respectively.

Spitzer observed NGC~3079 with the IRAC, IRS, and MIPS.
From the archived data, however, we find that the MIPS 160 $\mathrm{\mu m}$ and the IRAC 8 $\mathrm{\mu m}$ band images are heavily affected by saturation effects in the central region of NGC~3079.
Weedman et al. (2005) and Bernard-Salas et al. (2009) presented mid-IR spectra of the nuclear region between 5 and 37 $\mathrm{\mu m}$ obtained by the IRS.
These spectra showed that emissions from polycyclic aromatic hydrocarbons (PAHs) are dominant in the mid-IR.
The strong PAH emission suggests that the central activity of NGC~3079 is better characterized by a nuclear starburst rather than an AGN. 

In this paper, we present multi-band images of NGC~3079 in the near- to far-IR obtained by the AKARI satellite (Murakami et al. 2007).
AKARI has two instruments, the Infrared Camera (IRC; Onaka et al. 2007) and the Far-Infrared Surveyor (FIS; Kawada et al. 2007), with which we obtain 10 photometric band images at central wavelengths of 3, 4, 7, 11, 15, 24, 65, 90, 140, and 160 $\mathrm{\mu m}$ by pointed observations.
The 7 and 11 $\mathrm{\mu m}$ bands enable us to discuss the spatial distribution of the PAH emission.
The 4 far-IR bands are efficient to determine the properties of interstellar dust. 
We perform spatially-resolved studies of the interstellar dust in the whole galaxy of NGC~3079. 
Here we present non-saturated images as well as new maps at 11 $\mathrm{\mu m}$ and 15 $\mathrm{\mu m}$, although the spatial resolutions are slightly worse than Spitzer due to the smaller diameter of the AKARI telescope (68.5 cm; Kaneda et al. 2005; 2007).

\section{Observations and Data Reductions}

The IRC and FIS imaging observations of NGC~3079 were carried out in April and May, 2007.
They were performed in part of the AKARI mission program "ISM in our galaxy and Nearby Galaxies" (ISMGN; Kaneda et al. 2009).
A summary of the observation log is listed in table \ref{tab:log}.
With the IRC, we obtain $N3$ (reference wavelength of 3.2 $\mathrm{\mu m}$), $N4$ (4.1 $\mathrm{\mu m}$), $S7$ (7.0 $\mathrm{\mu m}$), $S11$ (11.0 $\mathrm{\mu m}$), $L15$ (15.0 $\mathrm{\mu m}$), and $L24$ (24.0 $\mathrm{\mu m}$) band images using a standard observation mode, IRC02; each image has a size of about $\timeform{10'} \times \timeform{10'}$.
The FWHMs of the Point Spread Functions (PSFs) are \timeform{4".0}, \timeform{4".2}, \timeform{5".1}, \timeform{4".8}, \timeform{5".7}, and \timeform{6".8} for the $N3$, $N4$, $S7$, $S11$, $L15$, and $L24$ bands, respectively (Onaka et al. 2007).
The near- and mid-IR images are created by using the IRC imaging pipeline software version 20071017 (Lorente et al. 2007). 

\begin{table*}
  \caption{Observation log.}\label{tab:log}
  \begin{center}
    \begin{tabular}{llll}
      \hline
      Date 			& Observation ID&Observation mode & Details \\
      \hline
      2007 Apr 30	&1401006		&FIS01			&Scan speed: \timeform{8"} $\mathrm{s^{-1}}$, Reset interval: 0.5 s \\
      2007 Apr 30	&1401005		&IRC02			&$N3$, $N4$, $S7$, and $S11$ band imaging\\
	  2007 May 1	&1401007		&IRC02			&$L15$ and $L24$ band imaging\\
      \hline
    \end{tabular}
  \end{center}
\end{table*}

For the IRC, we obtained both short and long exposure image data in a single observation.
In this paper, we mainly use long exposure images but the image in the $N3$, $N4$, $S7$, and $S11$ bands are affected by saturation near the nuclear region of the galaxy. 
The region affected by the saturation in the long exposure image is determined by visual inspection in comparison with the PSFs.
Then the pixels in the saturation-affected region plus 1-pixel (\timeform{2".3}) periphery are replaced by those unaffected in the short exposure image.
As a result, we have replaced a region of about \timeform{12"} $\times$ \timeform{12"} near the center in the $N3$, $N4$, $S7$, and $S11$ bands.

With the FIS, we obtain $N60$ (centered at a wavelength of 65 $\mathrm{\mu m}$), $WIDE$-$S$ (90 $\mathrm{\mu m}$), $WIDE$-$L$ (140 $\mathrm{\mu m}$), and $N160$ (160 $\mathrm{\mu m}$) band images using one of the FIS observation modes, FIS01.
It consists of two sets of round-trip slow scans with the user-defined parameters that are the scan speed of \timeform{8"} $\mathrm{s^{-1}}$, the cross-scan shift length of \timeform{70"}, and the reset time interval of 0.5 s.
A region of approximately $\timeform{10'}\times \timeform{20'}$ around the galaxy was covered with a scan map.
The FWHMs of the PSFs are $\sim \timeform{40"}$ for the $N60$ and $WIDE$-$S$ bands and $\sim \timeform{60"}$ for the $WIDE$-$L$ and $N160$ bands (Kawada et al. 2007) .
The FIS images are processed with the AKARI official pipeline software version 20070914 (Verdugo et al. 2007). 

\section{Results}

\begin{figure*}
  \begin{center}
    \FigureFile(160mm,){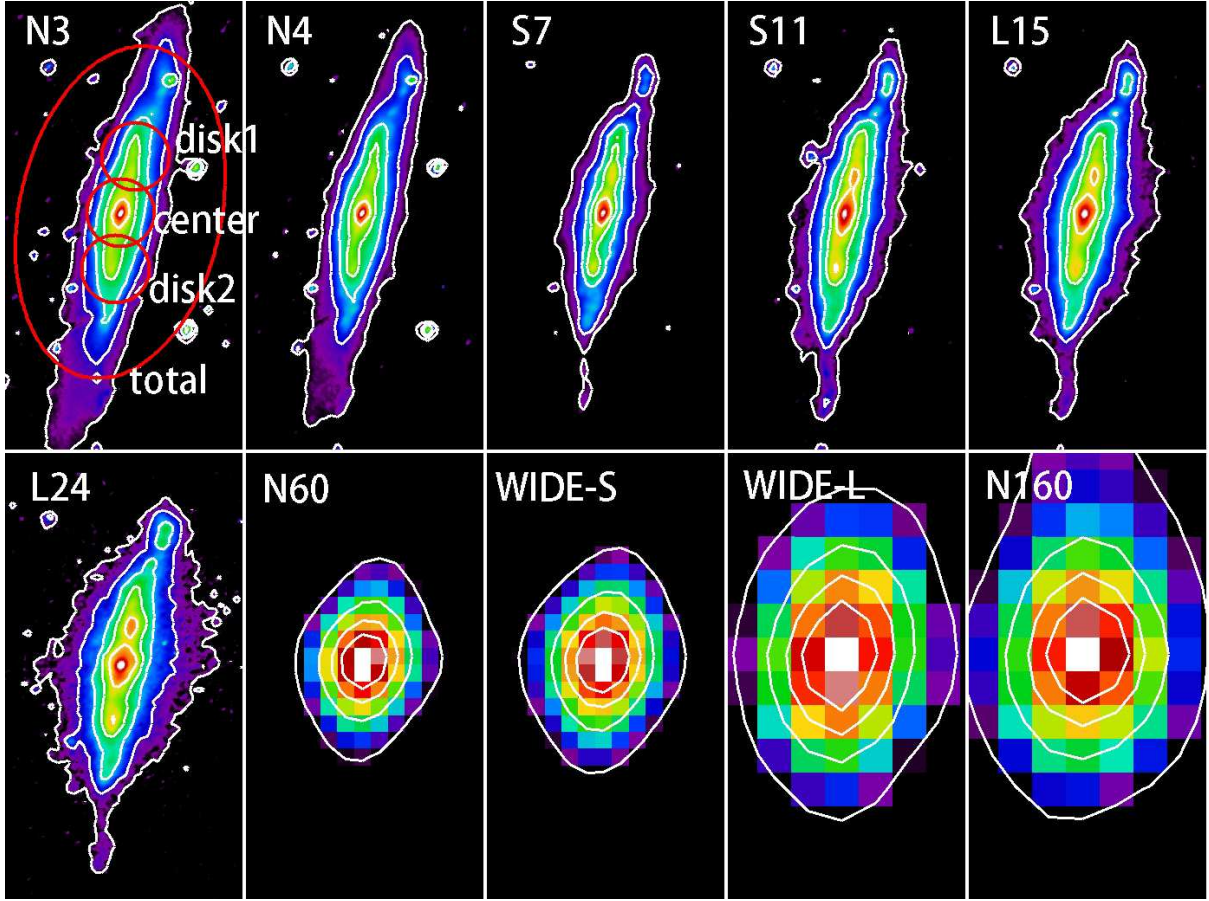}
  \end{center}
  \caption{AKARI 10 band images of NGC~3079 in a size of \timeform{6'.6} $\times$ \timeform{3'.5} for each. North is up and east is to the left. The contours are drawn at 0.1, 0.3, 1, 3, and 10 \% levels of the background-subtracted peak surface brightness for the near- to mid-IR images, and 10, 30, 50, and 70 \% for the far-IR images. The peak surface brightness is 91.4 $\mathrm{MJy~str^{-1}}$, 86.3 $\mathrm{MJy~str^{-1}}$, 631 $\mathrm{MJy~str^{-1}}$, 256 $\mathrm{MJy~str^{-1}}$, 296 $\mathrm{MJy~str^{-1}}$, 371 $\mathrm{MJy~str^{-1}}$, 194 $\mathrm{MJy~str^{-1}}$, 270 $\mathrm{MJy~str^{-1}}$, 166 $\mathrm{MJy~str^{-1}}$, and 85 $\mathrm{MJy~str^{-1}}$ for the $N3$, $N4$, $S7$, $S11$, $L15$, $L24$, $N60$, $WIDE$-$S$, $WIDE$-$L$, and $N160$ bands, respectively.}\label{fig:10bands}
\end{figure*}

The IR 10 band images obtained for NGC~3079 are shown in figure \ref{fig:10bands}.
The bin size is set to \timeform{0".5} for the near- to mid-IR bands, \timeform{15"} for the $N60$ and $WIDE$-$S$ bands and \timeform{30"} for the WIDE-L and $N160$ bands.
Smoothing by a gaussian kernel of each FWHM (Kawada et al. 2007) has been performed for the far-IR maps.
For every map, the image size is \timeform{6'.6} $\times$ \timeform{3'.5}, and the contours are drawn at 0.1, 0.3, 1, 3, and 10 \% of the background-subtracted peak surface brightness for the near- to mid-IR images, and 10, 30, 50, and 70 \% for the far-IR images.
The background levels are estimated by selecting several blank sky areas located around the galaxy, and taking median values of the surface brightness in these areas.
The IR emission from NGC~3079 is spatially resolved in all the bands including the far-IR bands.
The original images in the $N3$, $N4$, $S7$, and $S11$ bands show ghost signals at about \timeform{1'} in the northeast direction from the galactic center.
We have removed the ghost signals by subtracting scaled images of the galaxy from the original images, where a scaling factor is set to 3 \% for the $S11$ band and 0.5 \% for the others. 

As seen in figure \ref{fig:10bands}, the structure of the galaxy is different from band to band.
The near-IR images show more elongated distributions along the major axis of the optical disk than the mid- and far-IR images.
In the near-IR images, emission from evolved stars is prominent along the arms, while in the mid-IR images, the center and the arm regions are bright, among which relative intensities are different.
The mid-IR emission is attributed mostly to PAHs in the $S7$ and $S11$ bands (Weedman et al. 2005) and hot dust emission in the $L15$ and $L24$ bands.
In the far-IR images, cold dust emission from the galaxy is significantly extended along the edge-on disk.

Table \ref{tab:akarimatome} shows the flux densities of NGC~3079.
We set three circular aperture regions of \timeform{1'} in diameter. They are named disk1, disk2, and center as shown in figure \ref{fig:10bands}, along the disk of NGC~3079, where \timeform{1'} corresponds to $\sim$ 4 kpc for the distance to NGC~3079.
In addition, we obtain the total flux densities from the whole galaxy by integrating surface brightness within an elliptical aperture of \timeform{3'} and \timeform{5'} in minor and major axis, respectively, for the near- and the mid-IR images.
We use a circular aperture of \timeform{5'} in diameter for the far-IR images.
For the \timeform{1'} aperture that is not large enough as compared to the PSFs for the far-IR bands, we apply aperture corrections by using the correction table provided by Shirahata et al. (2009).
Color corrections are also applied with factors of 0.89--1.05 for the obtained flux densities of the far-IR bands, once we obtain a dust temperature.

\begin{table*}
  \caption{Flux densities of the 4 regions in NGC~3079.}\label{tab:akarimatome}
  \begin{center}
    \begin{tabular}{lllll}
      \hline
					&\multicolumn{4}{c}{Flux density (Jy)}\\
	  \hline
		Region\footnotemark[$*$]		&total				&center				&disk1				&disk2				\\
      \hline
		$N3$			&0.594 $\pm$ 0.015	&0.307 $\pm$ 0.007	&0.106 $\pm$ 0.003	&0.108 $\pm$ 0.003	\\
		$N4$			&0.424 $\pm$ 0.014	&0.231 $\pm$ 0.007	&0.075 $\pm$ 0.003	&0.074 $\pm$ 0.002	\\
		$S7$			&2.85 $\pm$ 0.07	&1.67 $\pm$ 0.04	&0.546 $\pm$ 0.013	&0.462 $\pm$ 0.011	\\
		$S11$			&2.13 $\pm$ 0.05	&1.10 $\pm$ 0.03	&0.440 $\pm$ 0.010	&0.407 $\pm$ 0.010	\\
		$L15$			&2.47 $\pm$ 0.06	&1.37 $\pm$ 0.04	&0.482 $\pm$ 0.014	&0.443 $\pm$ 0.012	\\
		$L24$			&3.62 $\pm$ 0.17	&1.95 $\pm$ 0.09	&0.718 $\pm$ 0.034	&0.601 $\pm$ 0.028	\\
		$N60$			&52 $\pm$ 10		&29 $\pm$ 5.8		&13 $\pm$ 2.6		&12 $\pm$ 2.5			\\
		$WIDE$-$S$		&103 $\pm$ 21		&54 $\pm$ 11		&26 $\pm$ 5.3		&26 $\pm$	5.3 		\\
		$WIDE$-$L$		&120 $\pm$ 36		&53 $\pm$ 16		&38 $\pm$ 11		&37 $\pm$	11 			\\
		$N160$		&127 $\pm$ 38		&50 $\pm$ 15		&33 $\pm$ 9.8		&34 $\pm$	10 			\\
	  \hline
	  \multicolumn{5}{@{}l@{}}{\hbox to 0pt{\parbox{180mm}{
		\par\noindent
		\footnotemark[$*$] The regions are defined in figure \ref{fig:10bands}.
	  }\hss}}
    \end{tabular}
  \end{center}
\end{table*}

The fractional errors of the flux densities are estimated to be 2.3--4.7 \% for the near- and mid-IR bands by considering photon noise, fluctuation of the background levels, and uncertainties in the conversion factor of ADU to Jy (Tanab\'{e} et al. 2008).
The flux uncertainties including both systematic effects associated with the far-IR detectors and absolute uncertainties are estimated to be no more than 20 \% for $N60$ and $WIDE$-$S$ and 30 \% for $WIDE$-$L$ and $N160$ (Kawada et al. 2007; Verdugo et al. 2007).
In table \ref{tab:kako}, we summarize the total flux densities of NGC~3079 obtained from the previous mid-, far-IR, and submillimeter observations.
By comparing the flux densities in table \ref{tab:akarimatome} with those in table \ref{tab:kako}, we find that the total flux densities obtained by AKARI are consistent with those previously measured by IRAS and ISO within the above errors.

\begin{table}
  \caption{Summary of the total flux densities from the previous mid-, far-IR, and submillimeter observations of NGC~3079 (Young et al. 1989; Klaas \& Walker 2002; Braine et al. 1997; Stevens \& Gear 2000).}\label{tab:kako}
  \begin{center}
    \begin{tabular}{ccc}
      \hline
			& Wavelength ($\mathrm{\mu m}$)	&Flux density (Jy) \\
      \hline
      IRAS	&12	&2.8 $\pm$ 0.6\\
		&25	&3.5 $\pm$ 0.7\\
		&60	&53 $\pm$ 11\\
                &100    &97 $\pm$ 19\\
      ISO	&15	&1.9 $\pm$ 0.6\\
		&25	&6.2 $\pm$ 1.9\\
		&60	&44 $\pm$ 13\\
		&120    &136 $\pm$ 41\\
		&150    &124 $\pm$ 37\\
		&180    &100 $\pm$ 30\\
		&200    &72 $\pm$ 22\\
      SCUBA	&450    &7.9 $\pm$ 1.2\\
		&850    &1.1 $\pm$ 0.1\\
      IRAM	&1220   &0.35 $\pm$ 0.01\\     
	  \hline
    \end{tabular}
  \end{center}
\end{table}

\begin{figure*}
  \begin{center}
    \FigureFile(120mm,){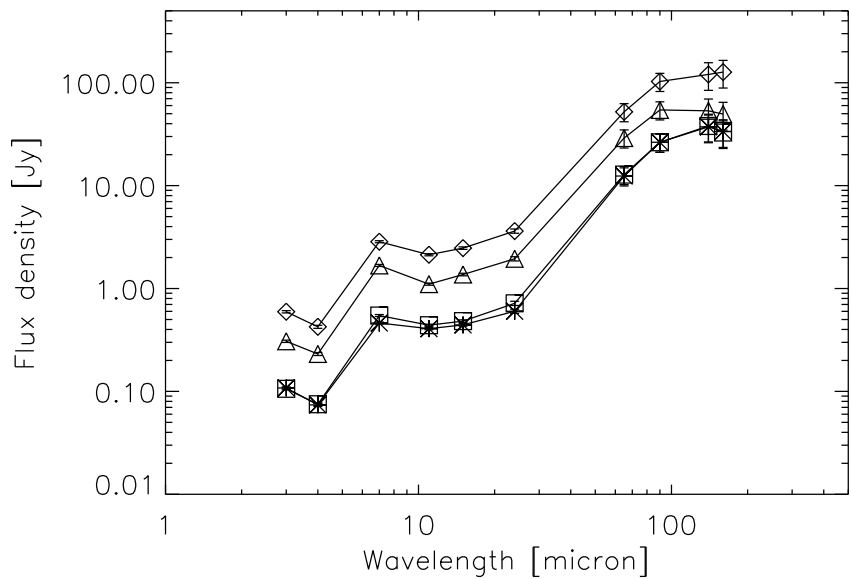}
  \end{center}
  \caption{SEDs of the total (diamond), center (triangle), disk1 (square), and disk2 (asterisk) regions created from the flux densities in table \ref{tab:akarimatome}. The error bar is attached to each data point.}\label{fig:sed}
\end{figure*}

\begin{figure*}
  \begin{center}
    \FigureFile(120mm,){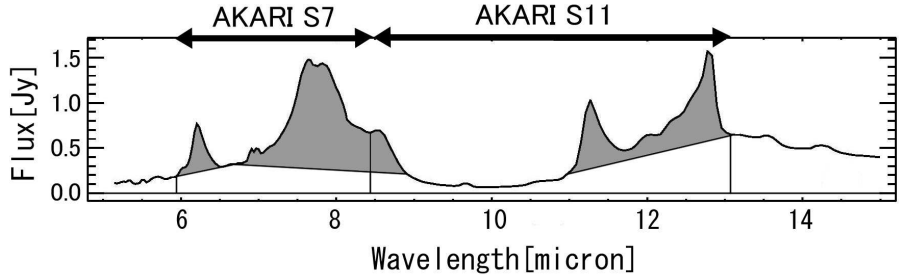}
  \end{center}
  \caption{Spectral coverages of the AKARI $S7$ and $S11$ bands (Onaka et al. 2007), overlaid on the Spitzer/IRS spectrum of the center of NGC~3079 taken from Weedman et al. (2005). We adopt the gray areas as contributions of emission from PAHs.}\label{fig:spitzer}
\end{figure*}

Figure \ref{fig:sed} shows the spectral energy distributions (SEDs) of the four regions constructed from the flux densities in table \ref{tab:akarimatome}.
Although these SEDs are quite similar from region to region as a whole, the $S7$ band intensity in the center region shows a relatively higher value than those in the other regions.
With the Spitzer/IRS, Weedman et al. (2005) showed that emissions from PAHs are dominant in the mid-IR, thus contributing much to the $S7$ and $S11$ band of AKARI.
Figure \ref{fig:spitzer} shows the spectral coverages of the AKARI $S7$ and $S11$ bands (Onaka et al. 2007) overlaid on the Spitzer/IRS spectrum of the center of NGC~3079 (Weedman et al. 2005).
From figure \ref{fig:spitzer}, we estimate that the PAH emission features contribute to at least $\sim$ 60 \% and $\sim$ 45 \% of the $S7$ and the $S11$ band intensity, respectively.
Thus the relatively high $S7$ band intensity in the center region indicates that the PAH emission from the $S7$ band is strong relative to that from the $S11$ band.
Allamandola et al. (1989) and Draine \& Li (2007) indicated that the PAH emissions of 6.2 and 7.7 $\mathrm{\mu m}$ are emitted by PAH cations in C-C stretching modes, while the emission of 11.3 $\mathrm{\mu m}$ is emitted by neutral PAHs in a C-H out-of-plane mode.
Neutral PAHs emit mcuh less in the 6--8 $\mathrm{\mu m}$ (Joblin et al. 1994; Kaneda et al. 2008).
Hence the ratio of the $S7$ to the $S11$ band intensity increases with the ionization degree of PAHs (Draine \& Li 2007).
The mid-IR SEDs in figure \ref{fig:sed} suggest that PAHs are highly ionized and radiation field is strong in the center of NGC~3079 as compared to those in the disk regions.
The far-IR SEDs in figure \ref{fig:sed} also indicate that a dust temperature is higher in the center than in the disk regions.
Hence the mid- and far-IR SEDs consistently suggest that star formation is relatively active in the central region of NGC~3079.

\section{Discussion}

\subsection{Dust Temperature and Mass}

\begin{figure*}
  \begin{center}
    \FigureFile(120mm,){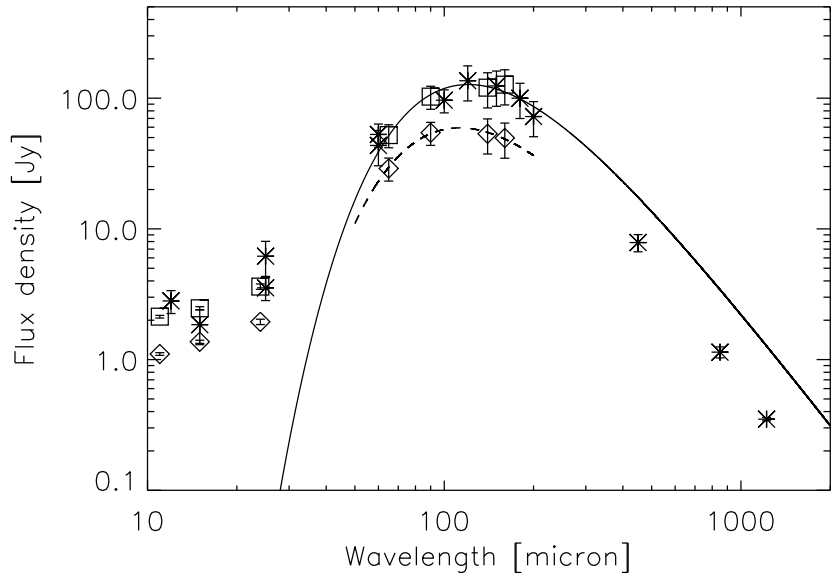}
  \end{center}
  \caption{Far-IR photometric data points (table \ref{tab:akarimatome} and \ref{tab:kako}) and best-fit graybody curves are plotted together. The squares indicate the total flux densities obtained by AKARI, while the asterisks indicate those from the previous observations (table \ref{tab:kako}). The diamonds show the flux densities obtained by AKARI for the 4 kpc center region. The solid line indicates the graybody curve ($\beta=1$) fitted to the far-IR (60--200 $\mathrm{\mu m}$) SED of the total region and extrapolated to the mid-IR and submillimeter regimes, while the dashed line indicates the graybody curve fitted to the far-IR SED of the center region.}
  \label{fig:sedfit}
\end{figure*}

\begin{figure*}
  \begin{center}
    \FigureFile(120mm,){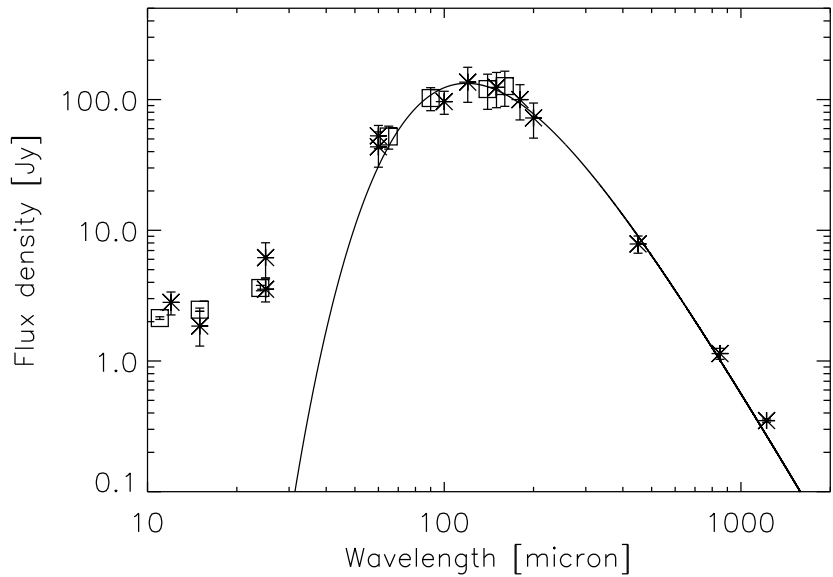}
  \end{center}
  \caption{Same as figure \ref{fig:sedfit}, but only photometric data points of the whole galaxy fitted by a graybody curve with $\beta =2$.}
  \label{fig:sedfit2+2}
\end{figure*}

We estimate dust temperatures by fitting the far-IR (60--200 $\mathrm{\mu m}$) SEDs of NGC~3079 by a single-temperature graybody model with an emissivity power-law index $\beta =1$.
For the total region of the galaxy, we use all the flux densities obtained from the previous observations (table \ref{tab:kako}) in addition to the AKARI/FIS results.
For the center, disk1, and disk2 regions, we use only the AKARI flux densities since the other far-IR observations could not spatially resolve the galaxy.
As a result, we derive the dust temperatures of $31\pm 1$ K, $32\pm 1$ K, $28\pm 1$ K and $28\pm 1$ K for the total, the center, the disk1, and the disk2 region, respectively (table \ref{tab:ratio}). 
The best-fit graybody curves for the total and the center region are shown in figure \ref{fig:sedfit}.
The curves reproduce the far-IR data points very well for both regions and we do not statistically need another component with a different temperature.
On the other hand, the model considerably overestimates the submillimeter 450, 850, and 1220  $\mathrm{\mu m}$ flux densities of the whole galaxy, which might reflect that the power-law index $\beta$ has a reasonable tendency to increase toward longer wavelengths (e.g. Hildebrand 1983).
In fact, from the submillimeter SED of NGC~3079, Braine et al. (1997) derived $\beta$ close to 2 between 100--1220 $\mathrm{\mu m}$.
Then, for comparison, we fit the far-IR SEDs by a graybody with $\beta=2$.
In figure \ref{fig:sedfit2+2}, we show the result for $\beta=2$, where we obtain poorer fits to the 60--200 $\mathrm{\mu m}$ SED, but with smaller discrepancies at longer wavelengths.

We calculate a dust mass by using the equation:
\begin{equation}
M_{\mathrm{dust}} = \frac{4a\rho D^2}{3}\frac{F_{\nu}}{Q_{\nu}B_{\nu}(T)}, 
\end{equation}
where $M_{\mathrm{dust}}$, $D$, $a$, and $\rho$ are the dust mass, the galaxy distance, the average grain radius, and the specific dust mass density, respectively. $F_{\nu}$, $Q_{\nu}$, and $B_{\nu}(T)$ are the observed flux density, the grain emissivity, and the value of the Planck function at the frequency of $\nu$ and the dust temperature of $T$.
We adopt the average grain radius of 0.1 $\mathrm{\mu m}$, and the specific dust mass density of 3 $\mathrm{g~cm^{-3}}$.
The grain emissivity factor is given by Hildebrand (1983) for $\beta=1$ and Alton et al. (1999) for $\beta=2$.
We use the 90 $\mathrm{\mu m}$ flux densities in table \ref{tab:akarimatome} and the temperatures derived from the above SED fitting for $\beta=1$ or $\beta=2$.
Since the optical depth at 90 $\mu$m is small ($\sim 2 \times 10^{-3}$) even in the center region, we detect far-IR emission from all the dust in each region. 

We show the dust masses thus derived for each region in table \ref{tab:ratio}. 
Similar dust masses are obtained between $\beta=1$ and $\beta=2$; the mass estimates for $\beta=1$ are slightly larger than those for $\beta=2$.

\subsection{Gas-to-dust mass ratio}

The total masses of atomic and molecular gas in table \ref{tab:ratio} are obtained from Irwin et al. (1987) and Devereux \& Young (1990), respectively.
However they used the old estimate of the distance to NGC~3079 (24.2 Mpc; Irwin et al. 1987).
For the new estimate (15.6 Mpc; Irwin \& Seaquist 1991, Sofue \& Irwin 1992), they correspond to $6.1 \times 10^{9} \MO$ and $6.0 \times 10^{9} \MO$, respectively.
The masses of atomic and molecular gas in the center in table \ref{tab:ratio} are obtained from Irwin \& Seaquist (1991) and Koda et al. (2002), respectively.
Koda et al. (2002) estimated the mass of molecular gas to be $5 \times 10^9 \MO$ within the central 2 kpc radius.
Thus, $\sim$ 80\% of molecular gas is concentrated in the central 2 kpc radius.
Since we use the same radius to define the center region in this paper, we can directly compare the dust mass with the molecular gas mass by Koda et al. (2002).
We estimate the H\emissiontype{I} mass in the same central region to be $1.2 \times 10^9 \MO$ from the H\emissiontype{I} radial profile in Irwin \& Seaquist (1991).

Then the gas-to-dust mass ratios are 860 for the total region and 1100 for the center region, which are considerably larger than the accepted value of 100--200 for our Galaxy (Sodroski et al. 1997).
The gas-to-dust mass ratio of similarly famous nearby edge-on starburst galaxy NGC~253 that is known to have starburst activity in the nucleus is $\sim 150$ for the central 4 kpc region and $\sim 70$ for the whole galaxy (Radovich et al. 2001; Kaneda et al. 2010).
These facts suggest that NGC~3079 has unusually high gas-to-dust mass ratios.

Klaas \& Walker (2002) derived the dust-to-gas mass ratio of 200 for the whole galaxy of NGC~3079 from the ISO observation, which is apparently inconsistent with our result.
Their result was based on the total molecular mass of $2.2 \times 10^9 \MO$ from Braine et al. (1997), which was estimated by assuming that the dust emission is proportional to the hydrogen column density.
Hence Klaas \& Walker (2002) adopted the dust-based molecular gas mass.
On the other hand, we use a larger molecular gas mass of $5 \times 10^9 \MO$ from the detailed CO mapping observation of the central 4 kpc region by Koda et al. (2002), which explains the above inconsistency.

We evaluate $L_{\mathrm{IR}}$, which is the total luminosity of the dust emission integrated between 1--1000 $\mathrm{\mu m}$ (table \ref{tab:ratio}) and derive the star formation rate (SFR) of NGC~3079 using the following equation (Kennicutt 1998):
\begin{equation}
\mathrm{SFR} = 1.7 \times 10^{-10} \frac{L_{\mathrm{IR}}}{\LO}~[\MO~\mathrm{yr^{-1}}].
\end{equation}
As a result, the SFR of NGC~3079 is 5.6 $\MO~\mathrm{yr^{-1}}$ and 2.6 $\MO~\mathrm{yr^{-1}}$ for the whole galaxy and the center 4 kpc region, respectively.
These SFRs are not high, slightly larger than that of our Galaxy.
The SFRs of NGC~253 were estimated to be 3.3 $\MO~\mathrm{yr^{-1}}$ and 2.1 $\MO~\mathrm{yr^{-1}}$ for the whole galaxy and the center 6 kpc region, respectively (Radovich et al. 2001).
Hence NGC~3079 and NGC~253 show similarly low values of the SFRs.
Engelbracht et al. (1998) concluded that NGC~253 is in a late phase of the starburst, having passed a rapid decrease of the star formation rate.
In contrast, from the above unusually high gas-to-dust mass ratios of NGC~3079, we suggest that NGC~3079 is still in an early phase of the starburst with a copious amount of dust-poor gas in the center, and active nuclear star formation is likely to occur in near future. 

\begin{table*}
  \caption{Derived properties of far-IR dust in NGC~3079.}\label{tab:ratio}
  \begin{center}
    \begin{tabular}{cccccc}
      \hline
		Region\footnotemark[$*$]		&total		&center		&disk1	&disk2		\\
      \hline
		$T_{\mathrm{dust}} (\beta=1)$	&31$\pm$ 1 K	&33 $\pm$ 1 K	&28 $\pm$ 1 K	&28 $\pm$ 1 K	\\
		$M_{\mathrm{dust}} (\beta=1)$	&$1.4 \times 10^7 \MO$	&$5.6 \times 10^6 \MO$	&$6.5 \times 10^6 \MO$	&$6.5 \times 10^6 \MO$ \\
		$T_{\mathrm{dust}} (\beta=2)$	&25$\pm$ 1 K	&26 $\pm$ 1 K	&23$\pm$ 1 K &23$\pm$ 1 K \\
		$M_{\mathrm{dust}} (\beta=2)$	&$1.2 \times 10^7 \MO$	&$4.8 \times 10^6 \MO$	&$5.2 \times 10^6 \MO$ &$5.2 \times 10^6 \MO$ \\
		$M_{\mathrm{gas}}~\mathrm{H\emissiontype{I}}$	&$6.1 \times 10^{9} \MO$	&$1.2 \times 10^{9} \MO$	&\dots	&\dots  \\
		$M_{\mathrm{gas}}~\mathrm{H_{2}}$	&$6.0 \times 10^{9} \MO$	&$5 \times 10^9 \MO$	&\dots	&\dots  \\
		$M_{\mathrm{gas}}/M_{\mathrm{dust}} (\beta=1)$	&860	&1100	&\dots	&\dots \\
		$L_{\mathrm{IR}}$	&$3.3 \times 10^{10} \LO$  &$1.5 \times 10^{10} \LO$ &\dots	&\dots \\
		SFR	&$5.6~\MO~\mathrm{yr^{-1}}$ &$2.6~\MO~\mathrm{yr^{-1}}$ &\dots	&\dots \\
	  \hline
	  \multicolumn{5}{@{}l@{}}{\hbox to 0pt{\parbox{180mm}{
		\par\noindent
		\footnotemark[$*$] The regions are defined in figure \ref{fig:10bands}.
	  }\hss}}
    \end{tabular}
  \end{center}
\end{table*}

\subsection{Spatial variations of PAH emission features}

\begin{figure}
  \begin{center}
    \FigureFile(60mm,){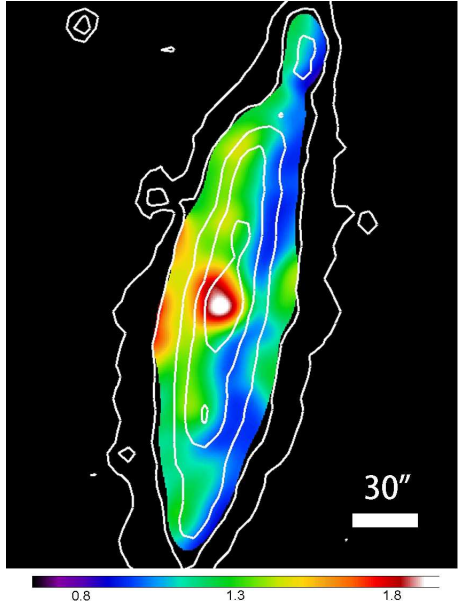}
  \end{center}
  \caption{Ratio map of the $S7$ to the $S11$ band surface brightness, superposed on the same $S11$ contour map as figure \ref{fig:10bands}.}\label{fig:hi}
\end{figure}

Now we focus on the mid-IR bands, the $S7$ and $S11$ bands that reflect the physical states of PAHs; as shown in figure \ref{fig:spitzer}, the $S7$ band contains the PAH 6.2 $\mathrm{\mu m}$ and 7.7 $\mathrm{\mu m}$ emission features, while the $S11$ band includes the PAH 11.3 $\mathrm{\mu m}$ and 12.7 $\mathrm{\mu m}$ emission features.
The images in figure \ref{fig:10bands} suggest that PAHs in the disk emit in the $S11$ band strongly relative to the $S7$ band as compared to PAHs in the center.  
Then we calculate the ratios of the $S7$ to the $S11$ surface brightness distribution after smoothing each image by a gaussian kernel of \timeform{10"} in FWHM and mask the area in the $S7$ image with brightness levels lower than 1.2 $\mathrm{MJy~str^{-1}}$ that corresponds to 0.6 \% of the peak brightness after smoothing.
The result is shown in figure \ref{fig:hi}.
The position of the central peak of the galaxy is adjusted within the accuracy of \timeform{0".5} between each image.
Figure \ref{fig:hi} reveals two prominent characteristics; a high $S7$/$S11$ ratio in the nucleus and asymmetry in $S7$/$S11$ ratio with respect to the disk.
The high $S7$/$S11$ ratio in the nucleus supports that ionized PAHs are dominant and thus radiation field is strong in the nucleus.
Along the disk, there are at least three spots showing relatively high $S7$/$S11$ ratios, which are located at the edge of the arms. The high spots located at the diagonal positions by \timeform{30"} off the center are probably due to artifacts originating from the PSFs (i.e. diffraction patterns of the telescope).

The ratio is systematically high on the eastern side of the galaxy with respect to the edge-on disk.
This asymmetry might be attributed to the extinction by extraplanar dust, especially silicate grains that have absorption features around 10 $\mathrm{\mu m}$, i.e. larger extinction in the $S11$ band and thus higher $S7$/$S11$ ratios on the eastern side of the galaxy.
This situation is consistent with the known configuration of NGC~3079 that the galaxy is highly inclined and the eastern side of the galaxy is far from us (Koda et al. 2002).

\subsection{Relation with nuclear activity}

\begin{figure*}
  \begin{center}
    \FigureFile(160mm,){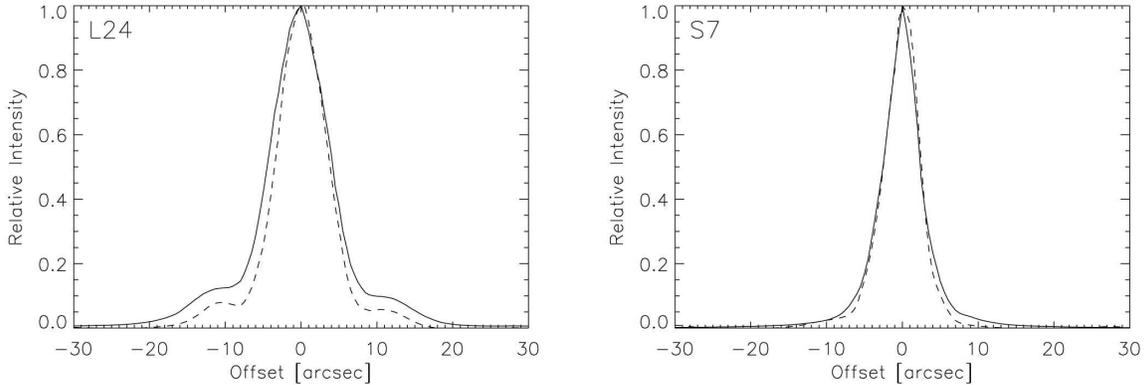}
  \end{center}
  \caption{Cutting profiles of the central emissions of NGC~3079 in the $L24$ (left) and the $S7$ band (right) along the minor axis of the galaxy (solid line), compared to the PSF (dash line).}\label{fig:47tuc}
\end{figure*}

The $L24$ image in figure \ref{fig:10bands} shows an extended structure in the center.
The structure is likely related to the activity and kilo-parsec-scale outflows of the galactic center, which were reported previously in H$\alpha$ and X-rays (e.g. Cecil et al. 2001).
We compare the central image of NGC~3079 to the PSF in the $L24$ band that is obtained from an isolated star in a globular cluster 47 Tuc (observation ID: 1700001, 1700002) by evaluating the cutting profile along the minor axis of NGC~3079 with a width of \timeform{2".5} (figure \ref{fig:47tuc}).
The images are normalized to the peak values and aligned in the instrumental pixel coordinates.
Figure \ref{fig:47tuc} reveals that there is excess above the PSF up to \timeform{20"}, which corresponds to the physical size of $\sim$ 1.5 kpc, similar to the scale of the H$\alpha$/X-ray outflow.
The extended component occupies $\sim$20 \% of the $L24$ band total flux in the central 4 kpc region.
Since this size is by far larger than a typical size of a molecular torus around an AGN, 1-100 pc, the extended hot dust component seems to be not of an AGN torus origin.

We also show the cutting profile obtained for the $S7$ band image in figure \ref{fig:47tuc}.
As compared to the $L24$ band, excess above the outskirts of the PSF is quite small.
Similarly, the excess is relatively small in the $S11$ and $L15$ bands.
Hence the hot dust responsible for the $L24$ band emission is more extended than the PAHs responsible for the S7 band emission, possibly being entrained by the plasma outflow from the nucleus.
As for the far-IR dust, we cannot obtain a clear result whether or not the dust is extended along the minor axis due to relatively poor spatial resolutions

\section{Conclusions}
With AKARI, we obtain the near- to far-IR images of NGC~3079 in 10 photometric bands.
We derive SEDs for the 4 kpc center region, the two 4 kpc disk regions, and the whole galaxy.
The SEDs consist of far-IR continuum emission from dust with a single temperature of 28--33 K and strong mid-IR emission features from PAHs.
We derive the dust masses of $5.6 \times 10^6 \MO$ and $1.4 \times 10^7 \MO$ for the central 4 kpc region and the whole galaxy, respectively, and find that NGC~3079 has unusually high dust-to-gas mass ratios in the central 4 kpc region ($\sim$ 1100), and even for the whole galaxy ($\sim$ 860).
The ratio of the surface brightness distribution at the wavelength of 7 $\mathrm{\mu m}$ to that at 11 $\mathrm{\mu m}$ suggests spatial variations in the properties of PAHs; emission from ionized and neutral PAHs is relatively strong in the center and the disk regions, respectively. We detect extended emission from hot dust in the center, possibly associated with the superbubbles. Thus radiation field is strong and star formation is relatively active in the center. Yet the total infrared luminosities of the galaxy indicate rather low star formation rates of 2.6 $\MO~\mathrm{yr^{-1}}$ and 5.6 $\MO~\mathrm{yr^{-1}}$ for the central 4 kpc region and the whole galaxy, respectively. Hence we conclude that NGC~3079 is likely to be in an early phase of the starburst with a copious amount of dust-poor gas in the center region.

\bigskip

We would like to thank all the members of the AKARI project for their intensive efforts.
We also express many thanks to the anonymous referee for the useful comments.
This work is based on observations with AKARI, a JAXA project with the participation of ESA.


\end{document}